\documentclass[dvips]{article}

\usepackage{icrc2011}

\newcommand{\eh}[1]{\,\mathrm{#1}}

\newcommand{\ttt}[1]{\times10^{#1}}

\newcommand{\dg}{^{\circ}}

\newcommand{\degr}{$\dg$}

\newcommand{\pcnt}{$\eh{\%}$}

\newcommand{\mr}[1]{\mathrm{#1}}

\renewcommand{\epsilon}{\varepsilon}

\newcommand{\tin}[1]{_{\mr{#1}}}

\newcommand{\hessefs}{HESS~J1857+026}

\title{Mapping the extended TeV source \hessefs\   down to Fermi-LAT energies with the MAGIC telescopes}

\newcommand{\etal}{\MakeLowercase{\textit{et al. }}} 
\shorttitle{Klepser \etal Mapping \hessefs\  with MAGIC}

\authors{Stefan Klepser$^{1}$, Julian Krause$^{2}$,Michele Doro$^{3}$ for
the MAGIC collaboration}
\afiliations{$^1$IFAE, Edifici Cn., Campus UAB, E-08193 Bellaterra, Spain\\
$^2$Max-Planck-Institut f\"ur Physik, D-80805 M\"unchen,
Germany\\$^3$Universitat Aut\`onoma de Barcelona, E-08193 Bellaterra, Spain}
\email{klepser@ifae.es}

\abstract{\hessefs\  is an extended TeV gamma-ray source discovered by H.E.S.S.
very close to the Galactic plane. Located in the vicinity of the pulsar PSR J1856+0245, the
source represents a pulsar wind nebula (PWN) candidate. In Fermi-LAT data, 7 photons
above $100\eh{GeV}$ were associated to it as VHE J1857+0252 by Neronov and
Semikoz (2010), while in the previous MeV-GeV catalogs no associated source was reported yet. MAGIC was upgraded to a
stereoscopic Cherenkov telescope system in 2009, which substantially improved its
performance with respect to extended objects. We observed \hessefs\  in 2010 and
analysed 29 hours of good quality stereoscopic data, yielding a highly significant
detection. We present an energy spectrum from $100\eh{GeV}$ to $10\eh{TeV}$
and skymaps for two different energy regimes. The spectrum does not show
any indication for an inverse Compton turnover, while we find an 
intrinsic extension of $\sigma = 0.22\pm0.02\tin{stat}\pm0.02\tin{sys}$ in
the energy range of $200-1000\eh{GeV}$. We discuss the
possible PWN nature of the source and the performance of MAGIC with respect to extended sources.}
\keywords{MAGIC, pulsar wind nebulae, \hessefs, very-high energy gamma-rays}

\begin{document}
\maketitle

\section{Introduction}

The H.E.S.S. collaboration reported a series of unidentified
very high energy (VHE, $> 100\eh{GeV}$) gamma-ray objects found in their survey of
the Galactic plane \cite{hess}. Among them is the extended source \hessefs, which is
situated at only $0.056\dg$ from the Galactic plane
 and was estimated to have
a flux of about 16\pcnt\ c.u. (Crab units).
From $21\eh{h}$ of data and 223 excess events, H.E.S.S. derived a detection significance of
$8.7\eh{\sigma}$ and a spectrum was extracted from $800\eh{GeV}$ to
$45\eh{TeV}$ compatible with 
a power law of index $2.39\pm0.08$. The extension was slightly beyond the H.E.S.S. resolution,
$(0.11\pm0.08)\dg\times(0.08\pm0.03)\dg$, and a significant tail-like structure on
the northern edge of the object was denoted. The VERITAS team confirmed this
detection in 2009 from $13\eh{h}$ of data without providing more details
\cite{veritas}.

In 2008, after the H.E.S.S. detection, a Vela-like pulsar PSR J1856+0245 was found in close vicinity
of \hessefs\ \cite{arecibo}. It has a spin period of $81\eh{ms}$, a characteristic age of
$21\eh{kyr}$, and a spin-down luminosity of $4.6\ttt{36}\eh{ergs\,s^{-1}}$.
The authors of \cite{arecibo} suggest that the pulsar might be powering a pulsar wind nebula (PWN) that
produces VHE gamma rays through inverse Compton scattering. The distance to
the pulsar was found to be about
$9\eh{kpc}$, but could only be estimated from the dispersion measure, and is
therefore uncertain within a factor of 2-3. This large distance leads to the high spin-down flux, which
energetically makes it a viable candidate to feed the VHE object. A faint
X-ray emission detected by ASCA in the same area supported this idea (also
\cite{arecibo}).

On the MeV-GeV gamma-ray side, no source was reported in GeV neither by
the EGRET team nor in the first Fermi/LAT catalog. Only in late 2010, Neronov and Semikoz
\cite{neronov} stated a significant detection of 7 clustered photons above
$100\eh{GeV}$, and a corresponding flux of
$(20.3\pm7.4)\ttt{-10}\eh{cm^{-2}\,s^{-1}\,TeV^{-1}}$, normalized at
$100\eh{GeV}$. Even more recently, this detection was confirmed by the
Fermi/LAT team, stating a fitted position compatible with that found by
H.E.S.S. \cite{fermisympefs}. An indication for pulsation was not found, which is
conform with the idea that the gamma rays might emerge from inverse Compton
scattering of diffuse electrons in a PWN. Besides that,
no extension was
found, but no quantitative limit on the extension was given. On the same
conference, \hessefs\ appeared in a list of Fermi sources above $10\eh{GeV}$,
including an integral flux above $10\eh{GeV}$, and an index of $1.30\pm0.28$,
indicating a hardening at lower energies.

A way to
prove the PWN nature of \hessefs\ is to analyse the spectrum and the
morphology evolution with energy. Inverse Compton scattering in PWN is
expected to produce a curved spectrum with a peak that in many cases lies in
the TeV energy range (see e.g. \cite{hintonhofmann} for a review). Its position depends however on many factors and 
may be, as in case of the Crab Nebula, at energies as low as some tens of GeV. In the H.E.S.S. spectrum of \hessefs, no hint for a turnover
can be seen, indicating the peak energy must be at least below $800\eh{GeV}$. At the same time, low-energetic electrons have
a longer mean free path length, and might diffuse farther away from the
pulsar, so an extending morphology at lower energies might give another hint towards a PWN scenario.

The two MAGIC telescopes~\cite{magic_scineghe}, situated on the island of La Palma (28.8\degr~N, 17.8\degr~W,
$2220\eh{m\,a.s.l.}$), use the Imaging Atmospheric Cherenkov Technique (IACT)
to detect gamma rays above a threshold as low as $75-80\eh{GeV}$ in standard
trigger mode\footnote{defined as the peak of the true energy distribution after
all cuts and at low zenith angles}. Since it started operating in stereoscopic mode in summer 2009,
its background supression was substantially reduced and a
sensitivity\footnote{defined as source strength in c.u. needed to achieve $N\tin{ex}/\sqrt{N\tin{bkg}}=5$ in $50\eh{h}$ effective
on-time.}  of
$0.8\eh{\%\, c.u.}$ above $250\eh{GeV}$ could be achieved
\cite{icrc_performance}. Also, the MAGIC-II
camera comprises a bigger trigger area, which improved the off-axis
performance of MAGIC, a feature that is crucial for new source discoveries and extended sources.


\section{Data set}

We observed \hessefs\  on 33 days between 2010 July 11 and 2010 October
10, obtaining $29\eh{h}$ of good-quality data (effective on-time after all
cuts) at zenith angles between 25\degr and 36\degr. The data was taken in
wobble mode, pointing the telescopes at 4 different pairs of pointing directions
symmetric to the source position, to
achieve a better flatness of exposure. Two of these pairs were chosen to be at
0.4\degr distance from the source, embracing it like a cross, and two more at
0.5\degr, also like a cross. This setup allows a large variety of cross-checks
and performance studies between different wobble pairs.

\section{Analysis methods and performance}

The analysis of the data was done with the MARS analysis framework
\cite{mars}, including the latest standard routines for stereoscopic analysis, whose
performance is also presented on this conference \cite{icrc_performance}.
Besides the abovementioned sensitivity and analysis threshold, it provides an
angular resolution of about $0.055\dg$ ($0.075\dg$) at $1\eh{TeV}$
($250\eh{GeV}$). The energy estimator is calculated from the brightness of the
shower image, its reconstructed impact parameter and the zenith angle, using
look-up tables. For the gamma/hadron separation and gamma direction
estimation we use the random forest (RF) technique \cite{rf}. We tested both our
standard random forests (\textit{point-source} RF), which are trained with
gamma rays simulated as a point-source MonteCarlo (\textit{point-source} MC), and a
random forest trained with diffusely generated gammas (\textit{diffuse} RF/MC).

\subsection{Skymapping}

For the skymaps, a three-step algorithm is applied as reported in some detail
also on this conference \cite{icrc_advanced_analysis}. This algorithm models the background directly from the
data and in bins of azimuth, properly taking into account azimuthal
dependencies of the off-axis sensitivity. To facilitate the interpretation
of structures, the events are 
folded (smeared) with a Gaussian kernel in order to achieve a point spread
function with a Gaussian shape and a total width of $\sigma\tin{tot}^2= \sigma\tin{smear}^2 + \sigma\tin{analysis}^2$.
The width of the analysis-intrinsic resolution, $\sigma\tin{analysis}$, is
determined from MC, and cross-checked with Crab Nebula data. In the two skymaps
presented here, the smearing
kernel is adjusted such that $\sigma\tin{tot}=0.11\dg$. With our intrinsic
resolution of $0.055\dg - 0.075\dg$, a
finer smearing would be
possible to see more details, but the noise that comes with the higher trial
factor also distorts the image, so the $\sigma\tin{tot}$ we chose is a
compromise between low noise and high resolution.
%

\subsection{Spectral Analysis}

The spectrum was calculated for a circular area with a radius of 0.4\degr, centered on our fitted
position. To estimate the average effective area, a MC dataset was used with gammas
simulated on a ring around the observation center, with inner and outer radii
of $0.25\dg$ and $0.65\dg$. This is done to take the variations of acceptance
across the area of the source into account. Besides the tests described in the
next paragraph, we cross-checked the spectrum normalizing the background
estimation by event numbers in off-source areas, or by effective
on-time, and using 4 different unfolding algorithms~\cite{unfoldings}.

\subsection{Performance with Extended Sources}

We conducted several tests to evaluate the performance of our analysis tools
with respect to extended sources. The main conclusions we could draw are:

\begin{itemize}
\item The significance, skymap and spectral analyses done with
\textbf{point-source and diffuse RFs} agree well within statistical errors unless a very
tight cut in the \textit{hadronness} parameter is applied. A tight hadronness
cut is equivalent to a high dependency on MC, so this descrepancy is conform
with what can be expected from the method.
\item The spectrum shown in this contribution can be reproduced within
systematic erros even using our standard $0.4\dg$-offset \textbf{point-source MC instead of the ring-like
MC} described above.
\item Since the source peak is slightly misplaced with respect to the center
of observations, we have considerable geometric exposure inhomogeneities.
We therefore tried 2 different ways of \textbf{deriving a background
estimate}:
One is to simply use the background data from the anti-source position within each wobble sub-set, the
other one is taking it from the corresponding \textit{wobble partner} data set,
extracting it at the same relative focal plane coordinates as the source
position in the On-data. The latter should be subject to less systematic
errors, but both methods again turn out to agree within systematic errors.
\end{itemize}

Concluding, we see that the analysis scheme we opted for is very robust
against the possible variations of it. The results we present here are finally
derived using the diffuse RFs and MC, and the wobble partner background
estimation.

%

\section{Results}

\begin{figure*}[!t]
   \centerline{\includegraphics[width=3.in]{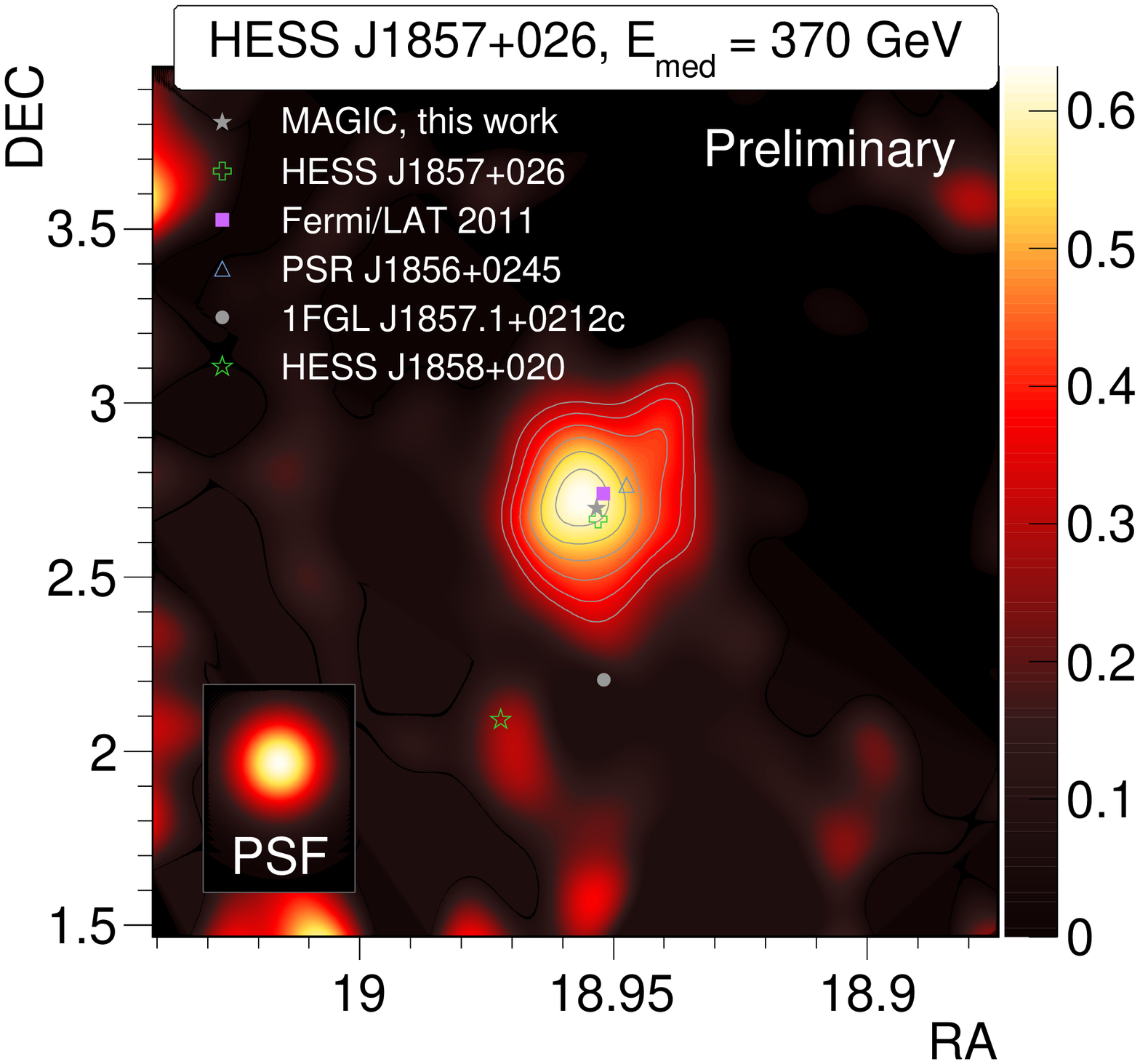}\label{fig2}
              \hfil
              \includegraphics[width=3.in]{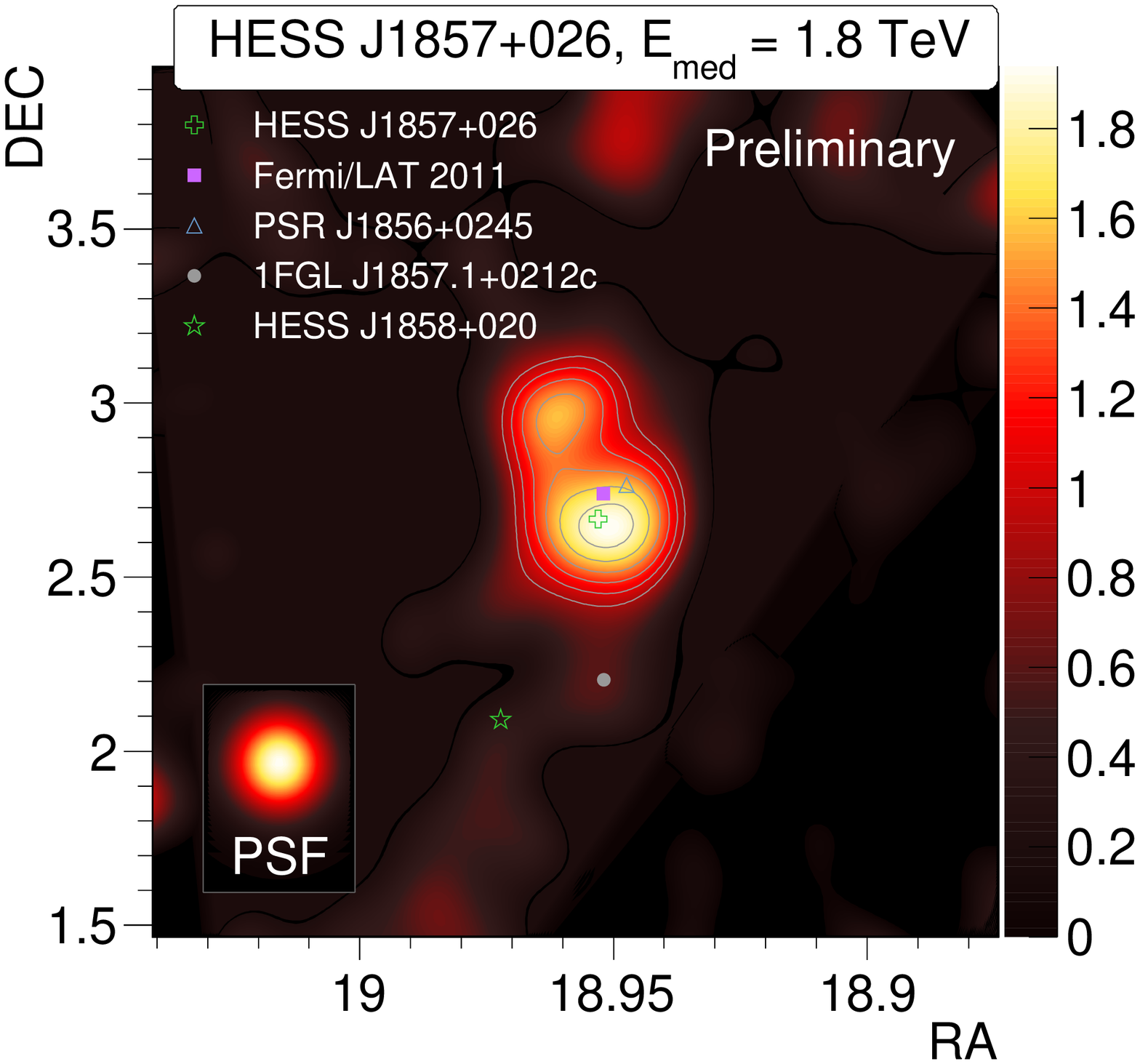} \label{fig3}
             }
   \caption{MAGIC "relative flux" map for events with
$200\eh{GeV}<E\tin{est}<1\eh{TeV}$ and $E\tin{est}>1\eh{TeV}$. The indicated
median energies are calculated from a MC that is reweighted to resemble the
spectrum of the source. The relative flux is calculated after smearing as $N\tin{ex}/N\tin{bkg,
<0.1\dg}$ and is roughly proportional to the actual gamma flux. Overlaid
are TS value contours in steps of 1, starting at 3. They roughly correspond to
Gaussian significances (for more information about
the relative flux and our TS, see \cite{icrc_advanced_analysis}). The fact
that we do not significantly detect HESS~J1858+20 might be attributed to the
MAGIC exposure at this distance to the observed sky positions, and will not be
discussed in this paper.
            }
   \label{figSkymaps}
 \end{figure*}

We detected $>3000$ gamma-like excess events above
$100\eh{GeV}$, providing us with a detection significance exceeding $12\eh{\sigma}$.
From these events, we calculated two skymaps for the energy range of $200\eh{GeV} - 1\eh{TeV}$, and
$>1\eh{TeV}$ (estimated energies). The median true energies of the gamma rays in these maps were
estimated from MC to be $370\eh{GeV}$ and $1.8\eh{TeV}$, respectively. The
results are
shown in Figure~\ref{figSkymaps}.

In the high energy map, there are two significant structures found, roughly
confirming what the H.E.S.S. team found. The energy regime in their map is probably
similar, although their higher effective area at high energies might lead to a
higher median energy.

The low energy skymap looks very different:
There is
only one structure, roughly centered at the position of \hessefs. We fit
the emission center to be located at
R.A.~$=(18.953\pm0.002\tin{stat}\pm0.002\tin{sys})\eh{h}$ and
Decl.~$=(2.70\pm0.02\tin{stat}\pm0.03\tin{sys})\dg$,
compatible within errors with the positions
determined by H.E.S.S. and Fermi/LAT.

In this low energy skymap, we find the source to be clearly extended beyond
the $\sigma\tin{tot}=0.11\dg$ implied by our PSF and smearing. The
source-intrinsic extension
we derive after subtraction of our total PSF in this energy range is
$0.22\dg\pm0.02\dg\tin{stat}\pm0.02\dg\tin{sys}$. This extension was fitted
assuming a two-dimensional Gaussian shape with or without circular symmetry
(the eccentricity is not significantly different from 0). It was also
cross-checked using different random forest matrices, as described in the
previous section.

\begin{figure}[!t]
  \vspace{5mm}
  \centering
  \includegraphics[width=3.in]{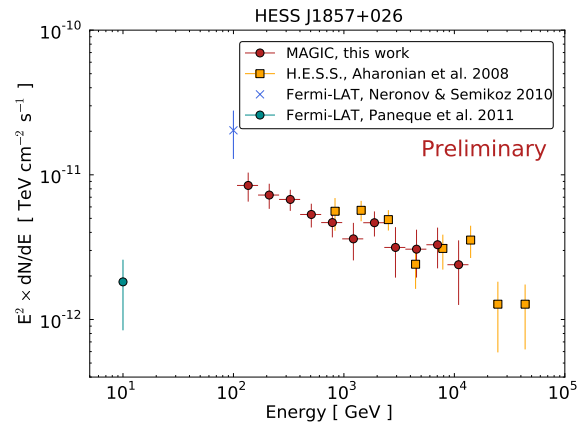}
  \caption{Spectral energy distributions of \hessefs\ measured by MAGIC,
H.E.S.S.~\cite{hess} and Fermi/LAT~\cite{neronov, fermisympdavid}. The MAGIC data
is unfolded to correct for migration and biasing effects, therefore the errors
are not independent.
}
  \label{figSpec}
 \end{figure}

The energy spectrum we derived
is shown in Figure~\ref{figSpec}. It spans over two orders of
magnitude and is compatible with a
power law of index $2.27\pm0.08\tin{stat}\pm0.1\tin{sys}$ and flux constant at
$1\eh{TeV}$ of
$(4.7\pm0.6\tin{stat}\pm1.4\tin{sys})\ttt{-12}\eh{TeV^{-1}\,cm^{-2}\,s^{-1}}$.
This is compatible with the parameters and spectral points found by
H.E.S.S., although maybe in slight disagreement with the Fermi/LAT estimate
given in
\cite{neronov}. The spectrum at and below $100\eh{GeV}$ is still being
investigated for systematic effects and will be published in a forthcoming
paper.

\section{Discussion and Outlook}

The extension we find in the energy regime of
$200\eh{GeV}-1\eh{TeV}$ is significantly higher than the extension reported by
H.E.S.S. at higher energies. This might support the idea that low-energetic
electrons from the PWN create this radiation after diffusing far away from the pulsar.
However, this reasoning only holds if the northern tail of the source is not
regarded part of the PWN, since it does not seem to have a low-energy
counterpart, implying a harder spectrum despite the larger distance to the
pulsar. The extension we see for the PWN at high
energies is still under investigation, but seems to be smaller than at low
energies (though still larger than the one
reported by H.E.S.S., which might be attributed to the possibly
higher median true energy of the H.E.S.S. detection and is not
necessarily a conflict).

The spectrum we find above $100\eh{GeV}$ shows no features nor a significant
curvature. This is not strongly contradicting a possible PWN scenario, but 
does not provide the evidence which a curved (IC motivated) shape would have
constituted. Adding the low integral flux and photon index reported in
\cite{fermisympdavid} (see Figure~\ref{figSpec}), and the fact that no source was previously found in the
MeV-GeV domain, allows for the conclusion that the spectrum has a very flat shape over 2-3
orders of magnitude, and a relatively hard turnover at
energies close to $100\eh{GeV}$. Compared to the well-studied PWN
HESS~J1825-137~\cite{hessetf, fermietf}, which is located at about half the
distance of \hessefs, but with a very similar pulsar age and period, we find the spectral parameters
and features are strikingly similar, except the TeV spectrum is more clearly curved in
HESS~J1825-137. 

Concerning the northern tail of the emission, it seems that it must be either
a different object altogether, or a different mechanism that is at work.
If it were connected to the main emission spot in some way, its distance to
the pulsar would be $\sim 50\eh{pc} \times
(d/9\eh{kpc})$, which does not exclude in general an association as for
instance through interactions of an expanding SNR shell from the same
supernova event. 
%

To possibly shed more light on the nature of this northern tail of the source,
we are presently working on a refined
analysis, trying to extract spacially resolved spectra, and more differential
skymaps. Also, we will reanalyse the data at and below $100\eh{GeV}$ to provide more
information about the position and shape of the spectral turnover.
Besides
that, we have to understand whether or not
there is a conflict with the Fermi/LAT analysis in \cite{fermisympefs}, which
claimed no extension at lower energies and might provide a quantitative
extension limit in the near future.

Concluding, we exploited the newly improved sensitivity of MAGIC to extended
sources to study the PWN \hessefs, and present a very wide, featureless energy spectrum and two
skymaps. We find that the northern, tail-like structure shown by H.E.S.S.
does not have a counterpart at lower
energies. Instead, only a more extended object can be seen at the position
compatible with the coordinates of \hessefs. We also find that the morphology and
overall spectral shape of the main
emission zone supports the concept of a PWN nature of \hessefs, which
therewith exhibits several similarities with the known PWN HESS~J1825-137. However, the wide-range, flat
TeV spectrum and the apparently different spectral behaviour of the northern
tail of the source might still indicate a more complex scenario for this
object. A more detailed analysis
of the MAGIC data, to be published in a forthcoming paper,
will provide a detailed spectral and morphological study that might
deliver more insight in the nature of \hessefs.

\clearpage

\end{document}